\documentclass{IEEEtran}

\usepackage{amsmath}
\usepackage{mathtools}
\usepackage{graphicx}
\usepackage{wrapfig}
\usepackage{caption}
\usepackage{booktabs}
\usepackage{array}

\begin{document}
\twocolumn[
\begin{@twocolumnfalse}

\title{Using Network Interbank Contagion in Bank Default Prediction}

\author{Riccardo Doyle*\\
*Royal Holloway, University of London
\\23 May, 2020\\}
\maketitle \thispagestyle{empty}

\begin{abstract}
Interbank contagion can theoretically exacerbate losses in a financial system and lead to additional cascade defaults during downturn. In this paper we produce default analysis using both regression and neural network models to verify whether interbank contagion offers any predictive explanatory power on default events. We predict defaults of U.S. domiciled commercial banks in the first quarter of 2010 using data from the preceding four quarters. A number of established predictors (such as Tier 1 Capital Ratio and Return on Equity) are included alongside contagion to gauge if the latter adds significance. Based on this methodology, we conclude that interbank contagion is extremely explanatory in default prediction, often outperforming more established metrics, in both regression and neural network models. These findings have sizeable implications for the future use of interbank contagion as a variable of interest for stress testing, bank issued bond valuation and wider bank default prediction.\\
\end{abstract}


\end{@twocolumnfalse}]

\section{Introduction}

In the aftermath of the Great Recession, the stability of established economic models came under scrutiny, leading to greater exploration of dynamic and macro-prudential methods that prioritize relationships between financial agents over their standalone characteristics. 

Among these is the study of interbank contagion, a subset of financial contagion seeking to quantify how the structure of linkages between banks' liabilities exacerbate loss propagation across the wider financial system.

Over the past decade, interbank contagion has been extensively covered in academic literature and widely adopted by regulators, with an extensive number of official bodies such as the IMF (Ozkan \& Unsal, 2012), the Bank of England (Bardoscia et. al, 2017) the Federal Reserve (Morrison et. al, 2017) and the European Central Bank (Hałaj \& Kok, 2013), including interbank contagion models in working papers and stress testing frameworks.

Despite widespread implementation and a rich history of academic discourse over the topic, there is little to no literature to examine whether or not interbank contagion models are accurate predictors of systemic risk and whether they are worth deriving conclusions from. Instead, studies thus far can be sub-grouped into either theoretical work concerning the network structure of the interbank market, or empirical work that only draws applied results from theoretical models, without ever verifying model accuracy in the first instance.

We believe this paper will address the aforementioned literature gap and make important contributions in determining whether or not interbank contagion models are explanatory and whether their implementation in regulatory contexts or scope for further research is justified.
 
To validate our earlier claims, let us offer an overview of existing literature with reference to methods and frameworks incorporated in this paper.

On a theoretical level, much effort has been placed in understanding how the structure of interbank networks influences the extent of contagion. Initial studies focused on foundational concepts, such as work from Allen and Gale (2000) showing that a complete network of interregional banks (where each is linked to every other) is more resilient to liquidity shocks than an incomplete one. Over time more complex analysis began to emerge. Nier et. al (2007) used non-empirical networks with ranges of key structure, probability and node size parameters to test contagion effects, finding – among other things – that the number of connections in a network can have dual effects on contagion. Initially, as connections increase, their ability to channel contagion increases alongside it; however, as connections increase past a certain large threshold, the newfound size of the network allows for greater risk sharing among nodes and decreases contagion considerably. Gai and Kapadia (2010) followed the same probabilistic and endogenous approach in Nier et. al (2007), but also introduced contagious effects originating from the sale of illiquid assets during crises. Much more importantly however, unlike previous literature, they differentiated between the probability and the extent of contagion. In doing so, they found that while the probability of contagion increases and subsequently decreases hyperbolically with network size, the extent of contagion spread increases until it reaches an asymptote, implying that while for large network sizes the probability of contagion is low, if contagion occurs it would produce extremely large repercussions on the system. 

In our paper, among other questions, we will seek to understand, in the context of the pre-financial crisis U.S. banking system, whether contagion is positively or inversely related to wider financial damage once a shock occurs.

Beyond theoretical studies on interbank market structure, several seminal studies have formalized the broad concepts of network structure in model form. The most widely adopted interbank contagion model is that of Eisenberg and Noe (2001), which proposes an iterative approach to network contagion, relating the default in payments of a borrowing bank to the devaluation in the assets of a lending bank, producing a clearing vector of payments that brings the banking system to equilibrium following successive periods of shock transmission. Another foundational model was that of Furfine (2003), which largely employs a similar contagion framework to Eisenberg and Noe (2001), but requires an exogenous trainable parameter for the recovery rate of payment defaults.
Model based papers of interbank contagion, compared to theoretical studies of market structure and empirical studies, are much less abundant. Since Eisenberg and Noe (2001) and Furfine (2003), the most meaningful new interbank contagion model advanced was that of Battiston et. al (2012), which proposes an innovative model (named DebtRank) based on the proportional propagation of contagion, in which banks would transmit shocks not through defaulted payments – as in previous models – but through gradual devaluation. They propose that a given interbank loan will lose value proportionally to the borrowing bank's loss in equity, transmitting contagion even before defaults occur.

In our paper, we have elected to adopt the DebtRank model framework, believing it to be more encompassing and flexible than alternative methods.

On an empirical level, many papers from the last decade have sought to apply the aforementioned theoretical studies. Upper and Worms (2004) produced an empirical network of interbank linkages simulating the German banking system, finding that in the worst-case scenario – and omitting safety mechanisms – the failure of a single bank could cause the banking system to lose up to 15\% of its total assets. Degryse and Nguyen (2007) took a more aggregate approach and, rather than focusing on a single event or time period, used time series data for the Belgian banking system between 1993 and 2002. They found that contagion risk in the system had changed considerably over time, increasing until 1997 and decreasing to a plateau after. They also drew conclusions on network structure, finding that Belgium's shift from a complete network akin to that of Allen and Gale (2000) to a more imperfectly connected system of select large players and numerous small players reduced contagion risk, arguing that the small banks could now no longer originate contagion. This assumes the large banks are fairly leveraged among each other, but as seen in the 2008 financial crisis, this is not always the case. Following a same national focus, Mistrulli (2011) tested for the extent of contagion in the Italian banking sector using unique data for the individual bilateral interbank assets and liabilities between each bank, which in most literature – including this study – are reconstructed using a maximum entropy framework. 
The paper adopted its own methodology for contagion transmission and applied a range of simulated shocks to draw conclusions on how capital ratios correlate with contagion, finding among other things that an increase in capital requirements does not stymie contagion.

A large number of successive empirical papers, such as (Upper, 2011), (Georgescu, 2015) , (Liu, 2020), (Gabrieli \& Salakhova, 2019) and (Leventides et. al, 2018), all apply real-world financial inputs from sample countries or banking systems to draw conclusions either on the extent of contagion potential in those systems or the relationship between financial indicators, network structure and contagion extent. 

No empirical study has tested the validity of interbank contagion as a field of study, instead accepting the models to be accurate ex ante and drawing policy conclusions given empirical inputs.

Additionally, due to the above, no study has also attempted to compare the explanatory power of interbank contagion with other systemic risk tools and predictors. 

This paper aims primarily to address these fundamental unanswered questions about interbank contagion and assess firstly whether it is at all effective as a systemic risk indicator, and secondly whether it is more or less effective than other leading systemic risk indicators.

\section{Methodology}

\subsection{DebtRank Contagion Transmission}

Many algorithms that model interbank contagion transmission are present in the literature. In this study we have elected to use DebtRank (Battiston et. al, 2012). Below, we will briefly explain the structure of the DebtRank algorithm. Let us define the assets and liabilities of every bank i in the system as the sum of its external and interbank components:

\begin{equation}
	A_i=\ A_i^{ext}+\ \sum_{j=0}^{n}{W_{ij}\ }\ \ \ 
\end{equation}

\begin{equation}
	L_i=\ L_i^{ext}+\ \sum_{i=0}^{n}{W_{ij}\ }\ \ \ \ 
\end{equation}

Where $W_{ij}$ is a matrix describing the loans (assets) extended by every bank i to every other bank j. Thus the sum of its rows, $\sum_{j=0}^{n}{W_{ij}\ }$, represents the total loans extended by bank i to all other banks in the system and constitutes the bank's interbank assets. 
$A_i^{ext}$ is a vector describing the total amount of external assets, defined as the total amount of assets excluding interbank assets $(A_i^{ext}=\ A_i-\ \sum_{j=0}^{n}W_{ij})$.

Consequently, let us define a bank's equity as the surplus of assets it holds:

\begin{equation}
	E_i=A_i-\ L_i\ \ \ \ \ 
\end{equation}

We now assume a mechanism for shock propagation that relates every bank to every other. We define a bank i's interbank assets in a given period as the interbank assets from the previous period devalued by the relative loss in equity of every other owing bank j: 

\begin{equation}
	W_{ij}^{t+1}=W_{ij}^t\frac{E_j^t\ }{E_j^{t-1}}\ \ \ \ \ 
\end{equation}

The logic behind the assumption is that as the equity of an owing bank j decreases, it will directly devalue the loan it received from bank i by that same proportional amount, as its ability to repay has decreased by $\frac{E_j^t\ }{E_j^{t-1}}$. 
We can re-write equation (4) in summative terms as:

\begin{equation}
\sum_{j=0}^{n}W_{ij}^{t+1}=(\sum_{j=0}^{n}W_{ij}^t\frac{E_j^t\ }{E_j^{t-1}})\ \ \ \ \
\end{equation}

Using this new form, we can then define the change in interbank asset value between consecutive periods as:

\begin{equation}
\sum_{j=0}^{n}{W_{ij}^{t+1}-}\sum_{j=0}^{n}W_{ij}^t=\sum_{j=0}^{n}W_{ij}^t\frac{E_j^t\ }{E_j^{t-1}}-\sum_{j=0}^{n}W_{ij}^t
\end{equation}

Or:

\begin{equation}
\sum_{j=0}^{n}{W_{ij}^{t+1}-}\sum_{j=0}^{n}W_{ij}^t=\sum_{j=0}^{n}{(W_{ij}^t}\frac{E_j^t\ }{E_j^{t-1}}-W_{ij}^t)
\end{equation}

Which can be rewritten to be:

\begin{equation}
\sum_{j=0}^{n}{W_{ij}^{t+1}-}\sum_{j=0}^{n}W_{ij}^t=\sum_{j=0}^{n}W_{ij}^t(\frac{E_j^t\ }{E_j^{t-1}}-1)\ \ \ 
\end{equation}

And further:

\begin{equation}
\sum_{j=0}^{n}{W_{ij}^{t+1}-}\sum_{j=0}^{n}W_{ij}^t=\sum_{j=0}^{n}W_{ij}^t\frac{{E_j^t-E}_j^{t-1}}{E_j^{t-1}}\
\end{equation}

Let us now assume that external assets and liabilities remain unchanged, which according to the original equity identity (3) only allows interbank assets to change. We can hence draw an equivalency between interbank asset changes and equity changes, as all else remains constant:

\begin{equation}
\sum_{j=0}^{n}{W_{ij}^{t+1}-}\sum_{j=0}^{n}W_{ij}^t=\ E_i^{t+1}-\ E_i^t\ 
\end{equation}

This allows us to re-write equation (9) as:

\begin{equation}
E_i^{t+1}-\ E_i^t=\sum_{j=0}^{n}W_{ij}^t\frac{{E_j^t-E}_j^{t-1}}{E_j^{t-1}}
\end{equation}

Or:

\begin{equation}
E_i^{t+1}-\ E_i^t=\sum_{j=0}^{n}\frac{W_{ij}^t}{E_j^{t-1}}({E_j^t-E}_j^{t-1})
\end{equation}

The term $\frac{W_{ij}^t}{E_j^{t-1}}$'s period t, is set to zero such that both the interbank assets of bank i and the change in bank j's equity between consecutive periods are relativized to their original values. At every period t the entire term is set to zero if a bank is insolvent or to itself otherwise:

\DeclarePairedDelimiter\Floor\lfloor\rfloor
\DeclarePairedDelimiter\Ceil\lceil\rceil

\[
  \varphi_{ij}^t=\
  \begin{cases}
                                   \frac{W_{ij}^0}{E_j^0} & \text{if $E_i\ \geq0$} \\
                                   0 & \text{if $E_i < 0$} \\
  \end{cases}
\]

We can then conclusively write the DebtRank model as:

\begin{equation}
E_i^{t+1}=max(0\ |\ E_i^t+\sum_{j=0}^{n}{\varphi_{ij}^t({E_j^t-E}_j^{t-1}))}
\end{equation}

At every iteration period, the model defines the equity of a bank, $E_i^{t+1}$, by removing the devaluation in its interbank assets, $\sum_{j=0}^{n}{\varphi_{ij}^t({E_j^t-E}_j^{t-1})}$, from the previous period's equity, $E_j^t$. If the devaluation is such that the bank's equity turns negative and the bank defaults, the model stores a value of zero instead.

The model iterates through consecutive periods until the relative difference in equity between consecutive periods is smaller than a certain threshold $\alpha$:

\begin{equation}
\frac{E_i^{t+1}-E_i^t}{E_i^t}<\alpha
\end{equation}

\subsection{DebtRank Linearity Coefficient}

In the previous section, DebtRank's shocks are propagated linearly according to the assumed propagation mechanism in equation (4). In this mechanism a percentage change in the equity of an owing bank j results in a corresponding percentage change in the interbank assets of a receiving bank i, implying the relationship between the two is linear and one to one.
A modification to the model can be applied by the insertion of a coefficient $\beta$ in equation (14):

\begin{equation}
E_i^{t+1}=max(0\ |\ E_i^t+\sum_{j=0}^{n}{\varphi_{ij}^t\beta({E_j^t-E}_j^{t-1})})
\end{equation}

Such that a percentage change in the equity of an owing bank j now results in a corresponding $\beta$ percent change in the interbank assets of a receiving bank i, since:

\begin{equation}
\footnotesize
\varphi_{ij}^t\beta\left({E_j^t-E}_j^{t-1}\right)=\ \frac{W_{ij}^t}{E_j^{t-1}}\beta{E_j^t-E}_j^{t-1}=W_{ij}^t\beta\frac{{E_j^t-E}_j^{t-1}}{E_j^{t-1}}
\end{equation}

This maintains the linearity in the proportion between one bank's equity and another's interbank assets but varies the gradient in the proportion. 

\subsection{Contagion Proxy}

The conventional output of the DebtRank algorithm is a solvency vector $\ \sigma_i$ describing whether each bank i is solvent or not. This constitutes a binary output, which limits contagion as a categorical variable. To produce a continuous variable with a richer information set, we define a proxy value for contagion as the percentage difference between the final equity state$\ E_i^{final}$ and the initial post-shock equity state $E_i^{post-shock}$:

\begin{equation}
Contagion\ Proxy=\ \frac{E_i^{final}-\ E_i^{post-shock}}{E_i^{post-shock}}100
\end{equation}

This metric produces an isolated measure of the equity lost by each bank i that is solely attributable to contagion, excluding the initial shock.

\subsection{Variable Setup}

The general testing framework consists in the formulation of established explanatory X attributes in bank default prediction, namely:

\begin{itemize}

\item Return on Assets (ROA).

\item Return on Equity (ROE). 

\item Past Due Short-Term Loan Book Value on Total Assets (Short Term Bad Loans as a percentage of Assets).

\item TIER 1 Capital Ratio. 

\item TIER 1 Leverage Capital Ratio. 

\end{itemize}

To these, we append an additional explanatory X attribute representing the output of the previously outlined DebtRank algorithm, hereafter referred to as the:

\begin{itemize}

\item Contagion Proxy.

\end{itemize}

All X attributes are tied to a binary dependant variable Y (taking 0 and 1 values) signifying whether each analysed bank has failed (0) or not (1). The high-level aim of the study is to verify for additional significance and explanatory power derived by the Contagion Proxy with respect to established variables.
To increase dimensionality, each attribute has 4 quarterly readings, resulting in 24 total X attributes.

\subsection{Method of Default Classification and Prediction}

Two machine learning classifiers will be used to predict the default Y outcome, given previously outlined X attributes. An overview of how each is applied is provided below.

\begin{center}
 {\ \emph{1. Neural Networks}}
\end{center}

Neural Networks have been widely applied in economics and finance with considerable success in a variety of use cases. They present advantages in their mapping of more complex, non-linear functions of X attributes, thus in this study they will be employed as a measure of hypothesized best classification performance. 
The shortcoming of neural networks however is their lack of transparency in variable significance. In this study we have elected to apply simple sensitivity analysis of the inputs through gradient descent to gauge explanatory power.

\begin{center}
 {\ \emph{1.1 Neural Network Setup}}
\end{center}

Several elements of the network have been kept constant. The network consists of three hidden layers, all of which have ReLU activations, and a single sigmoid activation (for binary classification) node as its output layer. Two dropout layers are present after the first and second hidden layer with a 10 percent dropout probability. Weights are initialized according to a random truncated normal distribution with mean 0 and standard deviation 0.2, while biases are initialized to 0.

\begin{center}
 {\ \emph{1.2 Neural Network Tuning}}
\end{center}

Data was split into training, validation and testing in equal thirds. All data was normalized using a Robust Scaler. The model trains on the training data, then selects the best combination of hyper-parameters based on accuracy performance in the validation set. The hyperparameters were combinations of:

Hidden Layer Structure $\in$ ([8, 16, 8], [4, 8, 16], [16, 8, 4]).

Solver  $\in$ (Stochastic Gradient Descent, Adam, RMSProp).

\indent Learning Rate $\in$ (0.001, 0.05, 0.1).

Once an ideal combination is identified, the model is re-trained using the tuned hyperparameters and evaluated Out-Of-Sample (OOS) on the testing dataset to obtain final performance.

\begin{center}
 {\ \emph{1.3 Neural Network Sensitivity Analysis}}
\end{center}

The model obtained in tuning represents our final, optimized neural network. It contains the fully backpropagated weights and biases that guaranteed it the lowest losses in binary cross entropy. We are now interested in the explanatory power of each input in the network. We can obtain an indication of this through gradient descent.
Let us provide a brief summary of the former and state key terms. 
A representation of our model, using a sample [16, 8, 4] node hidden layer architecture is presented in Fig. 1.

\begin{figure}[ht]
	\centering
	\captionsetup{justification=centering}
	\includegraphics[width=0.5\textwidth]{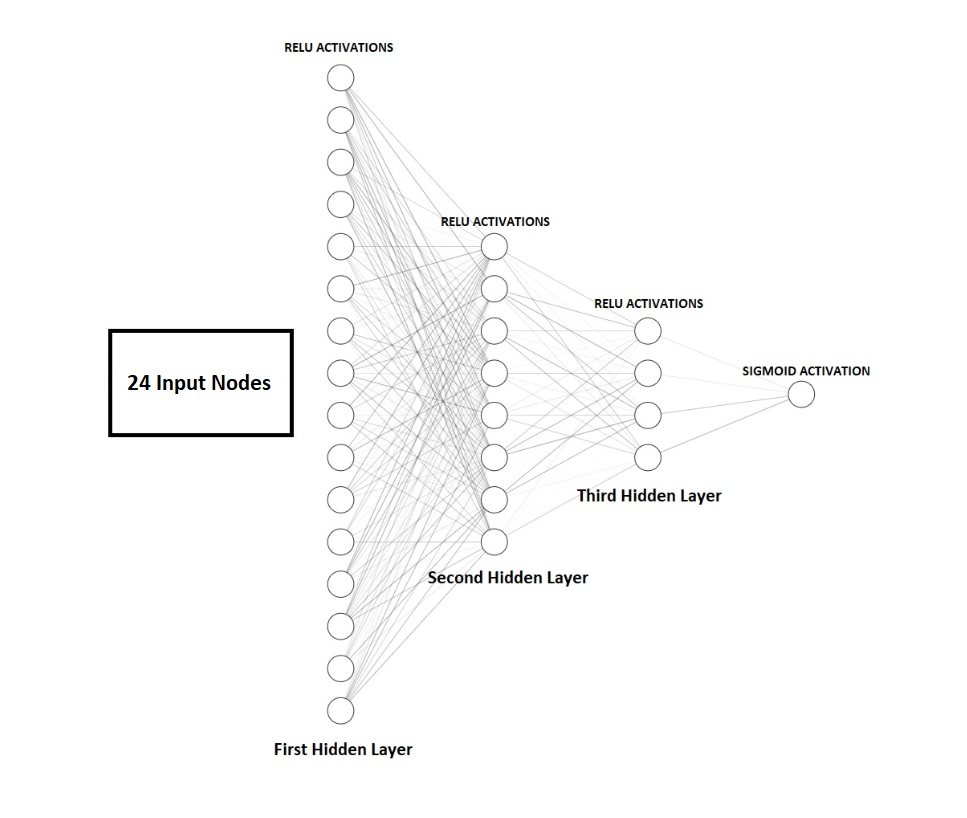}
	\caption{\emph{Sample neural network model.}}
\end{figure}

The model has an input layer consisting of 24 nodes, one for each explanatory X attribute we included in our dataset.
Let us denote the input layer, subsequent three hidden layers and output layer by the index j = 1:5. Let us further denote the number of nodes in each layer, starting from the top, by the index i = 1:n, where n is the last node.
Let us also define some key network components:

\begin{itemize}

\item 	$W_{i,k}^j$: The weight in layer j, connecting the i-th node in layer j to the k-th node in layer j+1.

\item $Z_{j,i}$: The input value at layer j that is fed into node i. This term is the sum of the product between each preceding layer's node and the weight that joins it to i. 

\item 	$A_{j,i}$: The activation function with $Z_{j,i}$ as input, such that $A_{j,i} = f(Z_{j,i})$, for some function f.

We can now more fully define $Z_{j,i}$ as:

\begin{equation}
Z_{j,i}=\ \sum_{i=1}^{n}{W_{i,k}^{j-1}\cdot\ A_{j-1,i}}
\end{equation}

\end{itemize}

Fig. 2 re-creates the network in Fig. 1 for just one input node who's effect on the model we want to assess.

\begin{figure}[ht]
	\centering
	\captionsetup{justification=centering}
	\includegraphics[width=0.46\textwidth]{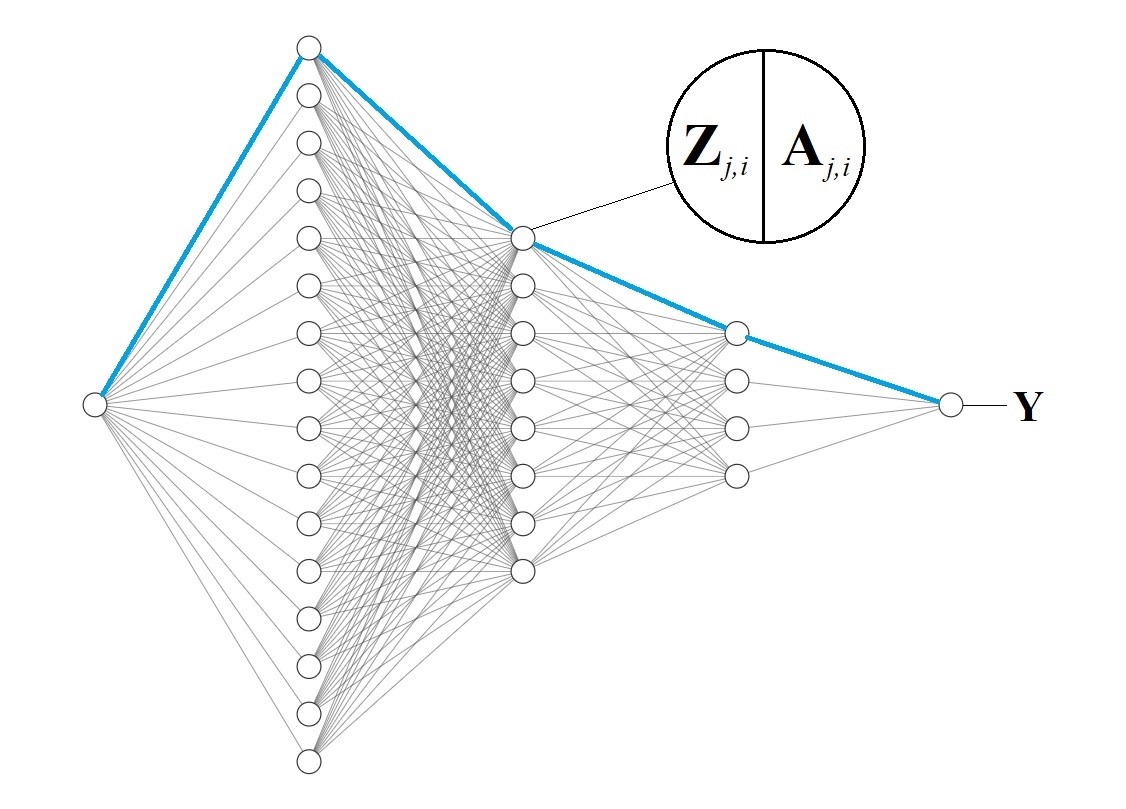}
	\caption{\emph{Sample backpropagation path (highlighted in blue) for a given forward pass from a single neuron to the next layer's single neuron through to output (Y).}}
\end{figure}

The marginal effect of this input node on the final output Y is determined by the marginal effect on each subsequent weight and node it is connected to.
To understand that general effect, let us begin with the local marginal effect on just one specific path in the network. We take the highlighted path in light blue from Fig. 2, going from the input node to the first node of each subsequent layer. Thus, we move from the input, or $A_{1,1}$, to $Z_{2,1}$ to $A_{2,1}$ to $Z_{3,1}$ … to the final sigmoid activation $A_{5,1}$ and the output Y.
The final output Y depends on the previous sigmoid activation $A_{5,1}$ with one-to-one transmission. Thus a unit change in $A_{5,1}$ will result in a unit change in Y, or in symbolic terms:

\begin{equation}
\frac{dY}{dA_{5,1}}=1
\end{equation}

Working backwards, the activation in the fifth layer, depends on its input in the same layer, formalized as:

\begin{equation}
\frac{dA_{5,1}}{dZ_{5,1}}
\end{equation}

Then by chain rule, we can say that the output depends on the aforementioned input according to:

\begin{equation}
\frac{dY}{dA_{5,1}}\cdot\frac{dA_{5,1}}{dZ_{5,1}}
\end{equation}

We can keep extending the chain rule backwards along the path until we arrive to the first input $A_{1,1}$:

\begin{equation}
\footnotesize
\frac{dY}{dA_{5,1}}\cdot\frac{dA_{5,1}}{dZ_{5,1}}\cdot\frac{dZ_{5,1}}{dA_{4,1}}\cdot\frac{dA_{4,1}}{dZ_{4,1}}\cdot\frac{dZ_{4,1}}{dA_{3,1}}\cdot\frac{dA_{3,1}}{dZ_{3,1}}\cdot\frac{dZ_{3,1}}{dA_{2,1}}\cdot\frac{dA_{2,1}}{dZ_{2,1}}\cdot\frac{dZ_{2,1}}{dA_{1,1}}
\end{equation}

Or:

\begin{equation}
\frac{dY}{dA_{5,1}}\ \cdot\ \prod_{j=5}^{2}{\ \frac{dA_{j,1}}{dZ_{j,1}}}\cdot\frac{dZ_{j,1}}{dA_{j-1,\ \ 1}}
\end{equation}

Which tells us how a unit change in the input, will affect the output Y for this particular path. To understand how the same unit change will affect the output Y across all possible paths generated by the input, we simply sum the gradients of each path:

\indent\\
\emph{for a given input node I}:
\indent\\

\quad $\frac{dY}{dA_{1,I}}=0$
\indent\\

\quad \emph{for a given second layer input node $\in$ (1, 2, 3 ... N)}:

\quad \quad \emph{for a given third layer input node $\in$ (1, 2, 3 ... M)}:

\quad \quad \quad \emph{for a given fourth layer input node $\in$ (1, 2, 3 ... O)}:

\indent\\

\quad \quad \quad $\frac{dY}{dA_{1,I}}=\ \frac{dY}{dA_{1,I}}+\ \frac{dY}{dA_{5,1}}\cdot\frac{dA_{5,1}}{dZ_{5,1}}\cdot\frac{dZ_{5,1}}{dA_{4,m}}\cdot\frac{dA_{4,m}}{dZ_{4,m}}\cdot\frac{dZ_{4,m}}{dA_{3,k}}\cdot$

\quad \quad \quad \quad \quad \quad \quad $\frac{dA_{3,k}}{dZ_{3,k}}\cdot\frac{dZ_{3,k}}{dA_{2,p}}\cdot\frac{dA_{2,p}}{dZ_{2,p}}\cdot\frac{dZ_{2,p}}{dA_{1,I}}$

\indent\\

Where N, M and O represent the number of nodes in the second, third and fourth layer respectively.

We will later report our findings according to this methodology, resulting in a single gradient for each input variable/node, quantifying how a unit change in the input $A_{1,I}$ affects the output Y.

To elaborate on the components above, the derivative of the input to a given node k in a given layer j with respect to the previous layer's j-1 activation from node i, $\frac{dZ_{j,k}}{dA_{j-1,i}}$, is given by the previous layer's weight from node i to node k, since:

\begin{equation}
\frac{d\left(\ \sum_{i=1}^{n}{W_{i,k}^{j-1}\cdot\ A_{j-1,i}}\right)}{dA_{j-1,i}}=\ W_{i,k}^{j-1}\ 
\end{equation}

The activation function derivatives with respect to the input, $\frac{dA_{j,k}}{dZ_{j,i}}$, are simply the gradients of the functions themselves. In our case, all hidden layers contain ReLU activations, which are defined as:

\[
  ReLU\left(Z_{j,i}\right)=\
  \begin{cases}
  0 & \text{for $Z_{j,i} < 0$} \\
  Z_{j,i} & \text{for $Z_{j,i}\ \geq0$} \\
  \end{cases}
\]

With a gradient of:

\[
  \frac{dA_{j,k}}{dZ_{j,i}}=\frac{d(ReLU\left(Z_{j,i}\right))}{dZ_{j,i}}=\
  \begin{cases}
  1 & \text{for $Z_{j,i} > 0$} \\
  0 & \text{for $Z_{j,i}\ \le0$} \\
  \end{cases}
\]

The final layer's activation is a sigmoid function, to smoothen outputs for binary classification. The sigmoid function is defined as:

\begin{equation}
Sigmoid\left(Z_{j,i}\right)=\ \frac{1}{1+e^{-Z_{j,i}}}
\end{equation}

With a gradient of:

\begin{equation}
\frac{dA_{j,k}}{dZ_{j,i}}=\frac{d(Sigmoid\left(Z_{j,i}\right))}{dZ_{j,i}}=\frac{e^{Z_{j,i}}}{\left(1+e^{Z_{j,i}}\right)^2}
\end{equation}

Results obtained through these gradients will reveal the sensitivity of the output to each of the inputs in the study.

\begin{center}
 {\ \emph{2. Logistic Regression}}
\end{center}

Logistic regression models are used extensively in financial literature concerning default prediction and can be viewed as a reliable baseline for this type of analysis. We employ them due to their greater transparency in both coefficient and significance analysis. 

Logistic models are non-linear and follow a log of the odds specification:

\begin{equation}
log\frac{P(X)}{1-P(X)}=\ \sum_{j=0}^{K}{b_jx_j}
\end{equation}

Where, P(X) is the probability of default, $b_j$ is the regression's j-th coefficient and $x_j$ is the regression's j-th variable.
Exponentiating both sides to the base of e results in a more interpretable specification of:

\begin{equation}
\frac{P(X)}{1-P(X)}=\ e^{\sum_{j=0}^{K}{b_jx_j}}
\end{equation}

Or:

\begin{equation}
\frac{P(X)}{1-P(X)}=\prod_{j=0}^{K}e^{b_jx_j}\
\end{equation}

Where the left-hand side of the equation represents the odds of P(X) occurring (the probability of X happening over it not happening), while the right-hand side is the product of the exponentiation of each regressor.
The interpretation of the model is thus that a one unit change in the value of some variable $x_j$ will result in an $e^{b_j}$ fold increase in the odds of P(X) occurring.

Our use of logistic regressions will be two-layered in this study:

Our particular use of logistic regression will also involve the application of a Lasso penalty to observe (a) if contagion remains among the significant variables (b) the magnitude and direction of its coefficients. 

\subsection{Testing Framework}

We begin the study by selecting a period of observation during which we want to record attributes. This period is the crash from the financial crisis and the first stages of its recovery, i.e. all four quarters from 2009. We do this with a view of predicting defaults occurring right after the aforementioned period. It would be reasonable to predict defaults from immediately after or during the crash, however the entire notion of interbank contagion is to identify the defaults of banks that don't immediately become insolvent during a crash, but might eventually become insolvent later as a result of prior immediate defaults. Hence we apply a lag that leads us to also include some recovery months as predictor periods.

In the stated 2009 window, we have attribute data on some 7,000 commercial banks in the U.S. financial system (only banks which remained solvent and as standalone entities for the entire period were included in this study). We then seek to identify which of these attributes caused certain banks to fail and others to remain solvent.

We ascertained we have a number of X attributes for these observations/banks over the specified period, but we require a Y label to discern whether they failed or not and whether the X attributes can predict such an event. As we will record X attribute data right up until the end of 2009, we will aim to predict bank defaults over the first quarter of 2010 (the successive data period). We used a separate dataset to retrospectively label all 7,000 banks in the 2009 yearly window as either still solvent in the first quarter of 2010 (1) or insolvent (0). The general framework of the study is to train a model on the X attributes in the 2009 window to predict the binary 2010 default state.
The problem is framed cross sectionally. The model is trained In-Sample (IS) on a subset of the banks (as opposed to using all banks but a subset of the overall time period) to optimize parameters and tested Out-Of-Sample (OOS) on a separate leftover subset of the banks. Overall predictive performance is observed on this leftover sample. 
The study will present findings on whether financial attributes, including contagion, up to a 1 year lag had any, and which, influence on default.

\begin{center}
 {\ \emph{Note on Class Balance}}
\end{center}

For each failed bank in our system there are roughly 50 non-failed banks, meaning our raw dataset is imbalanced and a classifier that labels all test samples as non-failed will be highly accurate without producing insight. To remedy this, we adopt a naïve random oversampling of the minority class (failed banks) with replacement until comparable to the majority one. To reduce computation times and to avoid excessive simplification of attributes in the oversampled class, we have reduced the overall dimensionality of the data to 1000 observations, evenly split between failed and non-failed institutions.

\begin{figure*}[t]
	\centering
	\captionsetup{justification=centering}
	\includegraphics[width=1\textwidth]{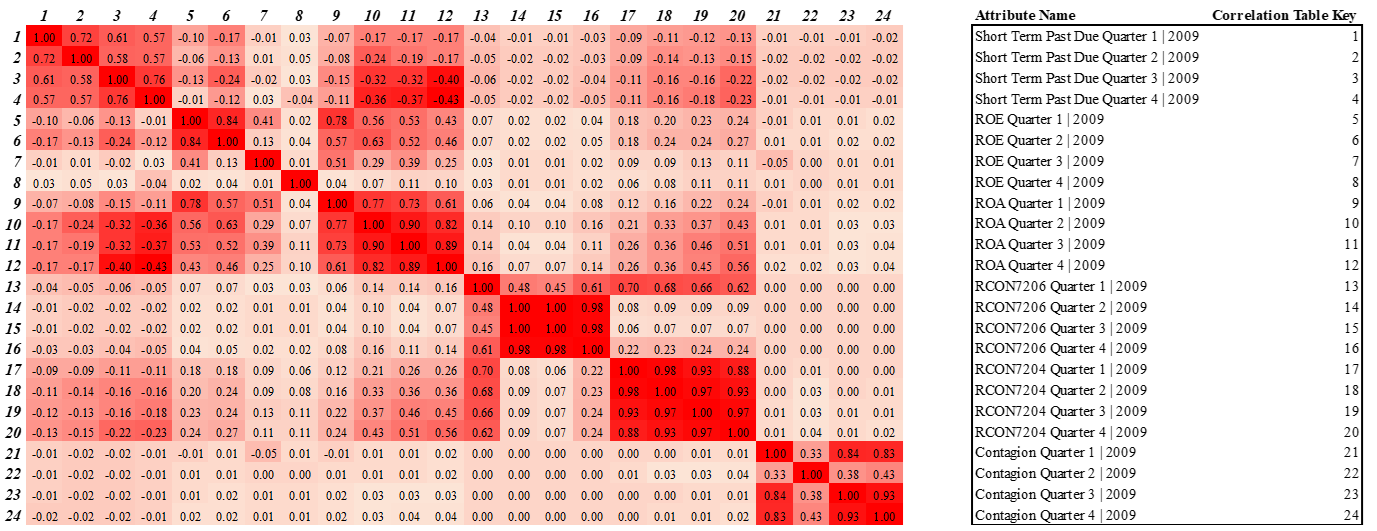}
	\caption{\emph{Correlation matrix of each attribute against every other. Attributes are substituted with number keys in the left table with a key dictionary table on the right. Coefficients in the correlation matrix are color coated according to a criterion that highlights high correlations in red, whether positive or negative, leaving more uncorrelated readings in  lighter hue.}}
\end{figure*}

\section{Data}

Data has been chiefly sourced from the Federal Financial Institutions Examination Council (FFIEC, 2019). The FFIEC provides datasets with balance sheet information on all United States commercial banks at quarterly intervals. 
Tailored code has been developed to compile and clean the semi-structured quarterly entries into a new, never before used, time-series dataset for each attribute mentioned in the “Variable Setup” section over a period of 2001-2018 (This constitutes a raw first entry of this study's X variables). A secondary source of data originates from the Federal Deposit Insurance Corporation (FDIC, 2019), which produces a list of all failed U.S. banks. The unique identifiers from the FDIC's dataset were matched to the FFIEC's, such that failed banks in the FFIEC dataset would take on a value of 1 and non-failed banks would take a value of 0. (This constitutes the “Y” dependant variable we would look to predict)
A data dictionary for variable abbreviations can be found at the Federal Reserve's Micro Data Reference Manual (Board of Governors of the Federal Reserve System, 2018).
We gathered all dataset variables necessary to either simulate or define the variables mentioned in the earlier model methodology. The data gathered was sufficient, but at times required manipulation. Interbank liabilities, for instance, had to be derived using related data, requiring the assumption of a closed interbank system.
Additionally, interbank data included information on the total interbank assets and liabilities of each bank, but not the interbank assets and liabilities of each bank with respect to every other, which is necessary to test contagion.

We hence applied a maximum entropy reconstruction algorithm (Anand et. al, 2015) capable of approximating the necessary data. At odd (31), n+1, and even (30), n, iteration intervals the algorithm normalizes the earlier introduced interbank asset matrix $W_{ij}$ to its total interbank assets (for iteration n) or liabilities (for n+1) to then multiply it by the observed total interbank assets (for n) or liabilities (for n+1) from the FFIEC dataset. In other words, this process rescales the rows and columns of the matrix to the dataset's aggregate values until a convergence is reached:

\begin{equation}
W_{ij}^n=\frac{W_{ij}^{n-1}}{\sum_{j}\ W_{ij}^{n-1}}{IA}_i^{FFIEC}\ 
\end{equation}

\begin{equation}
W_{ij}^{n+1}=\frac{W_{ij}^n}{\sum_{i}\ W_{ij}^n}{IL}_j^{FFIEC}
\end{equation}

The FFIEC data is used to iterate this paper's selected contagion models and its predicted defaults are compared with observed defaults from the Federal Deposit Insurance Corporation's Failed Bank List (FDIC, 2019) to generate conclusions.

\section{Results and Discussion}

\subsection{Preliminary Analysis of Variable Correlations}

Before analysing our model outputs, this section will cover some initial correlation analysis of studied attributes. Fig. 3 outlines correlation coefficients of each attribute in our study with respect to every other. 
We find generally that every attribute, with the exception of Return on Equity (keys 5 to 9), is highly autocorrelated across lags. In this regard, Return on Assets, Tier 1 Capital Ratio (RCON7206) and Tier 1 Leverage Capital Ratio (RCON7204) perform worst, while Contagion and Short Term Past Due loans perform moderately better but still considerably below Return on Equity.
In terms of cross attribute correlation, the picture changes. Predictably, Return on Equity is highly positively correlated with Return on Assets and moderately inversely related to Short Term Past Due loans. Short Term Past Due loans are strongly inversely correlated with Return on Assets and RCON7204. Return on Assets is highly positively correlated with Return on Equity and RCON7204. RCON7204 is highly positively correlated with Return on Assets and inversely correlated with Short Term Past Due loans. Finally, out of all attributes studied, Contagion and RCON7206 are the only two attributes to be largely uncorrelated with every other attribute, outside of its own lags, pointing to the high value they may bring into any analysis that involves these composite metrics. In particular, contagion is exceptionally uncorrelated, with no correlation coefficient having an absolute value magnitude greater than 0.05 excluding lag coefficients.
We summarize from this preliminary data analysis that contagion has promising qualities that may bring additional uncorrelated explanatory power to future regression analysis. We also conclude that some attributes have sizeable amounts of cross correlation with other attributes, and all attributes have a moderate to high degree of autocorrelation in their lags, hence dimensionality reduction methods will later be applied to avoid collinearity.

\subsection{Neural Network Results}

\begin{center}
 {\ \emph{Model Details}}
\end{center}

A neural network was tuned In-Sample and tested Out-Of-Sample for the purpose of variable analysis. We report the specifications of the final OOS tuned model in Table I.

\newcolumntype{P}[1]{>{\centering\arraybackslash}p{#1}}

\begin{table}[htbp]
  \centering
    \begin{tabular}{P{1.5cm}P{1.5cm}P{1.5cm}P{1.5cm}}
    \toprule
    \textbf{Hidden Layer Structure} & \textbf{Solver} & \multicolumn{1}{p{4.055em}}{\textbf{Learning Rate}} & \multicolumn{1}{p{4.055em}}{\textbf{OOS Accuracy}} \\
    \midrule
    (32, 16, 8) & RMSprop & 0.01  & 97.74\% \\
    \bottomrule
    \end{tabular}%
    \captionsetup{justification=centering}
    \caption{\emph{Classification accuracy and model specification of tuned OOS neural network model.}}
  \label{tab:addlabel}%
\end{table}%

The tuning resulted in an ideal learning rate of 0.01 combined with an RMSprop solver and a pyramid architecture of 32, 16 and 8 nodes over three hidden layers.
The model performed extremely well in default detection with 97.74 percent accuracy, meaning the vast majority of observations were correctly classified as either having remained solvent or failed in the first quarter of 2010. The high performance implies weights and magnitudes within the model are reliable indicators of how each variable affects default. 

\begin{center}
 {\ \emph{Sensitivity Analysis}}
\end{center}

As outlined in the methodology, we applied gradient descent to obtain derivatives of the network's output with respect to its inputs and gauge explanatory power. Table II reports our findings on a variable per variable basis.

\newcolumntype{P}[1]{>{\centering\arraybackslash}p{#1}}

\begin{table}[htbp]
  \centering
    \begin{tabular}{P{4.5cm}|P{3.2cm}}
    \toprule
    \multicolumn{1}{c}{\textbf{Variable/Attribute}} & \textbf{Output Gradient w.r.t Input: $\left(\frac{dY}{dA_{1,1}}\right)$} \\
    
    \midrule
    Short Term Past Due - Quarter 1, 2009 & 0.002 \\
    Short Term Past Due - Quarter 2, 2009 & -0.008 \\
    Short Term Past Due - Quarter 3, 2009 & 0.002 \\
    Short Term Past Due - Quarter 4, 2009 & 0.027 \\
    ROE - Quarter 1, 2009 & 0.041 \\
    ROE - Quarter 2, 2009 & -0.028 \\
    ROE - Quarter 3, 2009 & -0.018 \\
    ROE - Quarter 4, 2009 & -0.019 \\
    ROA - Quarter 1, 2009 & 0.006 \\
    ROA - Quarter 2, 2009 & 0.000 \\
    ROA - Quarter 3, 2009 & 0.002 \\
    ROA - Quarter 4, 2009 & -0.064 \\
    RCON7206 - Quarter 1, 2009 & 0.006 \\
    RCON7206 - Quarter 2, 2009 & -0.009 \\
    RCON7206 - Quarter 3, 2009 & -0.007 \\
    RCON7206 - Quarter 4, 2009 & -0.019 \\
    RCON7204 - Quarter 1, 2009 & 0.003 \\
    RCON7204 - Quarter 2, 2009 & -0.209 \\
    RCON7204 - Quarter 3, 2009 & -0.013 \\
    RCON7204 - Quarter 4, 2009 & -0.295 \\
    \textbf{Contagion Proxy - Quarter 1, 2009} & \textbf{-0.005} \\
    \textbf{Contagion Proxy - Quarter 2, 2009} & \textbf{-0.009} \\
    \textbf{Contagion Proxy - Quarter 3, 2009} & \textbf{0.002} \\
    \textbf{Contagion Proxy - Quarter 4, 2009} & \textbf{-0.029} \\
    \bottomrule
    \end{tabular}
  \captionsetup{justification=centering}
  \caption{\emph{Gradients of network's output value with respect to listed input variable.}}
  \label{tab:addlabel}%
\end{table}%

Before analysing the impact of contagion, let us generally describe the model and the overall structural integrity of the magnitude and direction of gradients, ensuring the model is broadly consistent with theory if it is to be used as a reliable benchmark for a novel attribute.

We would generally expect for Return on Equity (ROE) and Return on Assets (ROA) to be inversely correlated with default probability, meaning an increase in either results in lower chances of bankruptcy. The model broadly supports this, as three out of the four ROE entries have a negative gradient, signifying that an increase in ROE results in a decrease in the final 0 to 1 Y output closer to 0 where a solvent classification is more likely. ROA is more conflicting, as two of its readings have a negative gradient and two have a positive one, however the magnitude of its most significant reading (Quarter 4) is ten times larger than any of its other entries, signifying a net effect of inverse correlation with default. Short Term Past Due loans should be positively correlated with default, as a higher proportion of “bad loans” raises loss provisions, limiting banks' capital. This is confirmed by the data, with three of its four entries having positive gradients of much larger magnitude than its one negative reading. The two capital ratios, TIER 1 Leverage Capital Ratio (RCON7204) and TIER 1 Capital Ratio (RCON7206) should produce inverse correlations with default, as the metrics measure a bank's liquid capital buffer (where a higher ratio indicates a larger buffer). This is once again confirmed by the data as both have negative gradients for three out of four of their entries and insignificant magnitudes for their individual positive readings. Having ascertained the model is structurally consistent, let us analyse the magnitude and direction of contagion.

The Contagion Proxy would be expected to be inversely related to default, as it represents the equity lost due to contagion (note this value is negative). This means an increase in its reading mathematically represents a lower equity loss, thus a lower exposure to contagion and thus a lower default probability. We can confirm that this is the case, as three out of four contagion readings have a negative gradient and their magnitudes greatly overshadow that of its only positive gradient.

In addition to correct directionality, the Contagion Proxy also boasts extremely large gradients, where a higher gradient implies higher variable importance and impact on model output. Among variables with a negative gradient (15 variables) its reading for Quarter 4 of 2009 is the fourth largest by magnitude, coming after the second and fourth quarter readings of RCON7204 and the fourth quarter reading of Return on Assets. Most notably, the former contagion reading has larger gradients than any of the TIER 1 Capital Ratio (RCON7206) entries, representing confirmation within the context of this model that contagion can play a larger role in default prediction than long established “gold standard” predictors.


\newcolumntype{P}[1]{>{\centering\arraybackslash}p{#1}}

\begin{table*}[t]
  \centering
    \begin{tabular}{P{6cm}|P{5cm}|P{5cm}}
    \toprule
    \multicolumn{3}{c}{\textbf{OOS Accuracy: 93.7\%}} \\
    \midrule
    \multicolumn{1}{c}{\textbf{Variable/Attribute (Yearly)}} & \multicolumn{1}{c}{\textbf{Log of the Odds Coefficient}} & \textbf{Z Distribution P-Value} \\
    \midrule
    Short Term Past Due - Quarter 1, 2009 & \textit{\textbf{Lasso Reduced}} & \multicolumn{1}{c}{} \\
    Short Term Past Due - Quarter 2, 2009 & \textit{\textbf{Lasso Reduced}} & \multicolumn{1}{c}{} \\
    Short Term Past Due - Quarter 3, 2009 & \textit{\textbf{Lasso Reduced}} & \multicolumn{1}{c}{} \\
    Short Term Past Due - Quarter 4, 2009 & 0.0498 & \multicolumn{1}{c}{0.2671} \\
    ROE - Quarter 1, 2009 & \textit{\textbf{Lasso Reduced}} & \multicolumn{1}{c}{} \\
    ROE - Quarter 2, 2009 & -0.4107 & 0.0002* \\
    ROE - Quarter 3, 2009 & -0.0521 & \multicolumn{1}{c}{0.3879} \\
    ROE - Quarter 4, 2009 & -0.106 & 0.0290* \\
    ROA - Quarter 1, 2009 & \textit{\textbf{Lasso Reduced}} & \multicolumn{1}{c}{} \\
    ROA - Quarter 2, 2009 & \textit{\textbf{Lasso Reduced}} & \multicolumn{1}{c}{} \\
    ROA - Quarter 3, 2009 & \textit{\textbf{Lasso Reduced}} & \multicolumn{1}{c}{} \\
    ROA - Quarter 4, 2009 & \textit{\textbf{Lasso Reduced}} & \multicolumn{1}{c}{} \\
    RCON7206 - Quarter 1, 2009 & -0.0776 & \multicolumn{1}{c}{0.4007} \\
    RCON7206 - Quarter 2, 2009 & \textit{\textbf{Lasso Reduced}} & \multicolumn{1}{c}{} \\
    RCON7206 - Quarter 3, 2009 & \textit{\textbf{Lasso Reduced}} & \multicolumn{1}{c}{} \\
    RCON7206 - Quarter 4, 2009 & \textit{\textbf{Lasso Reduced}} & \multicolumn{1}{c}{} \\
    RCON7204 - Quarter 1, 2009 & \textit{\textbf{Lasso Reduced}} & \multicolumn{1}{c}{} \\
    RCON7204 - Quarter 2, 2009 & \textit{\textbf{Lasso Reduced}} & \multicolumn{1}{c}{} \\
    RCON7204 - Quarter 3, 2009 & \textit{\textbf{Lasso Reduced}} & \multicolumn{1}{c}{} \\
    RCON7204 - Quarter 4, 2009 & -1.8492 & 0.0000* \\
    \textbf{Contagion Proxy - Quarter 1, 2009} & \textit{\textbf{Lasso Reduced}} & \multicolumn{1}{c}{} \\
    \textbf{Contagion Proxy - Quarter 2, 2009} & \textit{\textbf{Lasso Reduced}} & \multicolumn{1}{c}{} \\
    \textbf{Contagion Proxy - Quarter 3, 2009} & \textit{\textbf{Lasso Reduced}} & \multicolumn{1}{c}{} \\
    \textbf{Contagion Proxy - Quarter 4, 2009} & \textbf{-0.1489} & 0.0018* \\
    \bottomrule
    \end{tabular}%
    
  		\captionsetup{justification=centering}
  		\caption{\emph{Coefficients from Logistic Regression following Lasso penalty. 24 original attributes reduced to 7 post-penalty.\\(*) Variable is significant at 95 percent confidence level.}}
  \label{tab:addlabel}%
\end{table*}%

\subsection{Logistic Regression Results}

A logistic model was fit to the study's 24 attributes alongside a Lasso regularization penalty to reduce dimensionality, co-linearity and increase significance of retained variables. 

Table III outlines our findings. The model's directionality is structurally consistent. Short Term Past Due loans are correctly directly proportional to odds of default, while Return on Equity, TIER 1 Capital Ratio (RCON7206), TIER 1 Leverage Capital Ratio (RCON7204) and the Contagion Proxy are correctly inversely proportional according to existing discussion in the Neural Network Results section of this study. Having established this, we may further analyse the model and the role of contagion in it. 

The applied Lasso regularization reduced the number of analysed attributes from 24 to 7. The substantial penalty implies that any retained variable can be assumed to be very impactful on default. Many attributes suffered heavy reductions, with all but Return on Equity (ROE) being limited to a single entry per metric. Return on Assets (ROA) has notably been discarded entirely. Contagion has been retained in the model with a single entry, offering an initial endorsement of its utility in the context of this model. 

To further elaborate, the Contagion Proxy's reading for the fourth quarter of 2009 yields a coefficient of -0.1489, which implies a one unit (standard normal intervals) increment in the Contagion Proxy would result in a $e^{-0.1489}\approx0.86$ factor multiplication (reduction) in the odds of default, ceteris paribus. Recall, as mentioned in the preceding Neural Network Results section, that a mathematical one unit  increase in the Contagion Proxy denotes a decrease in contagion exposure, given the proxy measures negative equity loss from contagion. The metric's coefficient is the third largest in absolute magnitude across the regression, ranking ahead of two of the three Return on Equity readings and the individual RCON7206 reading for same sign metrics. It is also extremely significant, with a Z-distributed p-value of 0.0018, representing the second most significant p-value in the regression, following only RCON7204's reading of $5.19\ \cdot\ 10^-16$.

\section{Conclusion}

This paper sought to explore the utility of interbank contagion as a default metric and more generally verify for its explanatory power and significance in the wider context of more established financial variables of interest. 
We adopted two default model frameworks, the first involving neural networks and the second involving logistic regression.
In the former, we concluded contagion has high explanatory power – with gradient magnitudes exceeding those of the TIER 1 Capital Ratio – and correct correlation with default – with greater contagion exposure being associated with greater odds of default.

In the logistic framework, we also concluded contagion to be highly explanatory – with a coefficient ranking third out of seven attributes by magnitude in a Lasso reduced model – and correctly correlated with default.
We hence summarize that not only contagion plays a clear role in bank default analysis, but that it is capable of even superseding the explanatory power and significance of far more established variables, justifying far greater interest in its analysis and application to stress testing, risk management and default prediction.

\section{Acknowledgements}

I would like to thank Dr. Fabio Caccioli for providing helpful comments.

\section{Competing Interests Statement}

I confirm there have been no involvements or interests that might  question the integrity of this study or the opinions stated therein.

\end{document}